\documentclass[12pt]{iopart}
\usepackage{graphicx}
\usepackage{amsfonts}
\usepackage{url}
\usepackage{epstopdf}
\usepackage{color}
\usepackage{tikz-cd}
\usepackage{bm}
\usepackage{setstack}

\def\L{\mathcal L}

\def\pa{\partial\Omega}
\def\E{{\mathbb E}}
\def\P{{\mathbb P}}
\def\R{{\mathbb R}}

\def\Ai{\mathrm{Ai}}
\def\Bi{\mathrm{Bi}}

\def\X{\bm{X}}

\def\D{{\bf D}}
\def\K{{\bf K}}

\def\x{\bm{x}}

\def\dD{{\epsilon}}  

\begin{document}

\title[A unifying approach to first-passage time distributions...]
{A unifying approach to first-passage time distributions in diffusing
diffusivity and switching diffusion models}

\author{Denis~S.~Grebenkov}
  \ead{denis.grebenkov@polytechnique.edu}
\address{
Laboratoire de Physique de la Mati\`{e}re Condens\'{e}e (UMR 7643), \\ 
CNRS -- Ecole Polytechnique, University Paris-Saclay, 91128 Palaiseau, France}

\date{\today}

\begin{abstract}
We propose a unifying theoretical framework for the analysis of
first-passage time distributions in two important classes of
stochastic processes in which the diffusivity of a particle evolves
randomly in time.  In the first class of ``diffusing diffusivity''
models, the diffusivity changes continuously via a prescribed
stochastic equation.  In turn, the diffusivity switches randomly
between discrete values in the second class of ``switching diffusion''
models.  For both cases, we quantify the impact of the diffusivity
dynamics onto the first-passage time distribution of a particle via
the moment-generating function of the integrated diffusivity.  We
provide general formulas and some explicit solutions for some
particular cases of practical interest.
\end{abstract}

\pacs{02.50.-r, 05.40.-a, 02.70.Rr, 05.10.Gg}



\noindent{\it Keywords\/}: diffusing diffusivity; switching diffusion; 
escape problem; first-passage time; diffusion-limited reaction

\submitto{\JPA}

\maketitle

\section{Introduction}

An accurate description of biochemical reactions occuring in a
heterogeneous, dynamically re-arranging intracellular environment is a
long-standing problem
\cite{Luby-Phelps00,Loverdo08,Benichou10,Barkai12,Benichou14,He16}.
Various theoretical aspects of the underlying intracellular transport
have been intensively studied over the past twenty years
\cite{Bressloff13,Hofling13}.  In particular, different theoretical
models have been proposed to account for molecular caging in the
overcrowded cytoplasm
\cite{Bouchaud90,Metzler00,Sokolov12,Metzler14}, viscoelastic
properties of the cytoskeleton polymer network
\cite{Guigas07,Szymanski09,Weber10,Goychuk12,Bertseva12,Grebenkov13a},
structural organization inside the cell \cite{Sadegh17}, random
diffusivity \cite{Manzo15}, and intermittent character of the motion
\cite{Benichou11,Bressloff17}.  All these mechanisms affect the
dynamics of molecules inside the cell, determine the statistics of
their first-passage times (FPT) to the binding sites, and thus control
the associated biochemical reactions.

Recently, we proposed a theoretical framework for investigating
diffusion-limited reactions in dynamic heterogeneous media
\cite{Lanoiselee18b}.  Modeling the effect of a rapidly re-arranging
medium as random changes of the amplitude of thermal fluctuations felt
locally by the tracer is based on the concept of diffusing diffusivity
introduced by Chubynsky and Slater \cite{Chubynsky14} and later
explored by several authors
\cite{Jain16,Jain16b,Chechkin17,Lanoiselee18a,Sposini18}.  Here, the
diffusivity $D_t$ of the tracer is considered as a stochastic process,
which is independent of the tracer's position.  For this annealed
model, we derived a general expansion for the propagator
$P(\x,t|\x_0,D_0)$ of the tracer, i.e., the probability density of
finding the tracer, started from $\x_0$ at time $0$ with the initial
diffusivity $D_0$, in a vicinity of a point $\x$ at time $t$:
\begin{equation} \label{eq:propagator}
P(\x,t|\x_0,D_0) = \sum\limits_{n=1}^\infty u_n^*(\x_0) \, u_n(\x) \, \Upsilon(t; \lambda_n|D_0) ,
\end{equation}
where the sum runs over all eigenvalues $\lambda_n$ and
$L_2$-normalized eigenfunctions $u_n(\x)$ of the Laplace operator
$\Delta$ in a confined bounded medium $\Omega \subset \R^d$, $\Delta
u_n + \lambda_n u_n = 0$, with mixed Dirichlet/Neumann boundary
conditions on the boundary $\pa$ of $\Omega$ \cite{Lanoiselee18b}.  As
usual, the Dirichlet condition, $u_n(\x) = 0$ at $\x \in \Gamma
\subset \pa$, accounts for a perfectly reactive sink $\Gamma$ on the
boundary, whereas the Neumann condition $\partial_{\bm{n}_{\x}}
u_n(\x) = 0$ at $\x \in \pa\backslash\Gamma$ describes an inert
reflecting wall (an obstacle) on the remaining part of the boundary
(here $\partial_{\bm{n}_{\x}}$ is the normal derivative at the
boundary point $\x$ directed outward the domain).

While the structural and reactive properties of the medium are
captured via the Laplacian eigenfunctions and eigenvalues
\cite{Grebenkov13}, the function $\Upsilon(t;\lambda|D_0)$ introduces
the annealed disorder and couples it to the dynamics of the tracer.
The function $\Upsilon(t;\lambda|D_0)$ was shown to be the Laplace
transform of the probability density function of the integrated
diffusivity
\begin{equation}  \label{eq:Tt}
T_t = \int\limits_0^t dt' \, D_{t'}, 
\end{equation}
with the initial value $D_0$.  For homogeneous diffusion with a
constant diffusivity $D_0$, one gets $\Upsilon(t;\lambda) = \exp(-D_0
t\lambda)$ and retrieves the standard spectral expansion of the
propagator \cite{Gardiner,Redner}.  If the initial diffusivity $D_0$
is chosen randomly from a prescribed distribution (e.g., the
stationary distribution), the average of Eq. (\ref{eq:propagator})
yields the spectral expansion for the conventional propagator
$P(\x,t|\x_0)$
\begin{equation} \label{eq:propagator0}
P(\x,t|\x_0) = \sum\limits_{n=1}^\infty u_n^*(\x_0) \, u_n(\x) \, \Upsilon(t; \lambda_n) ,
\end{equation}
in which $\Upsilon(t; \lambda_n)$ is the average of
$\Upsilon(t;\lambda|D_0)$, see below.

From the expansion (\ref{eq:propagator}), one easily gets the survival
probability of the tracer, the macroscopic reaction rate and other
quantities of interest.  For instance, for a tracer started at $\x_0$,
the probability density of the first-passage time to the binding site
$\Gamma$ reads
\begin{equation}  \label{eq:rho}
\rho(t|\x_0) = - \sum\limits_{n=1}^\infty u_n^*(\x_0) \,  \partial_t \Upsilon(t; \lambda_n) \int\limits_\Omega d\x \, u_n(\x)  .
\end{equation}
The impact of dynamic heterogeneities onto the first-passage time
density is thus controlled by the function $\Upsilon(t;\lambda)$.
When the diffusing diffusivity is modeled by a Feller process
\cite{Feller51} (also known as the square root process or the
Cox-Ingersoll-Ross process \cite{Cox85}), an explicit form of the
function $\Upsilon(t;\lambda)$ was derived in \cite{Lanoiselee18a}
(see also Sec. \ref{sec:Feller}).  For this model, we obtained the
asymptotic behavior of the probability density $\rho(t|\x_0)$ and
showed how the annealed disorder broadens the first-passage time
distribution \cite{Lanoiselee18b}.

In the present paper, we further develop this theoretical approach by
considering a general form of the stochastic equation for the
diffusing diffusivity.  Relying on the Feynman-Kac formula for the
function $\Upsilon(t;\lambda|D_0)$, we build a general framework for
studying diffusion-limited reactions and the related first-passage
time problems in the realm of diffusing diffusivity models
(Sec. \ref{sec:diffdiff}).  In particular, we recall the main formulas
for the Feller process and we derive new ones for the reflected
Brownian motion on an interval.  Moreover, we present similar results
for another important class of models, in which a diffusing particle
randomly switches between states with different diffusivities
(Sec. \ref{sec:switching}).  Such switching diffusion models are often
employed to describe the dynamics in biological systems
\cite{Bressloff17,Sungkaworn17,Weron17,Yin10,Godec17}.  Quite
naturally, switching diffusion models appear as discretized versions
of diffusing diffusivity models.  We formalize this connection by
relating switching rates to the drift and volatility coefficients of
the stochastic equation determining $D_t$.  In this way, one gets a
computationally efficient way to access the statistics of the
first-passage time in both types of models.

\section{Diffusing diffusivity models}
\label{sec:diffdiff}

We consider a particle diffusing in a $d$-dimensional dynamic
heterogeneous environment, whose effect is modeled via a diffusing
diffusivity $D_t$ which obeys a general stochastic equation in the
It\^o convention:
\begin{equation}  \label{eq:dDt}
dD_t = \mu(D_t,t) dt + \sigma(D_t,t) dW_t ,  
\end{equation}
subject to the initial condition $D_{t_0} = D_0$, where $W_t$ is the
standard Wiener process, and functions $\mu(D,t)$ and $\sigma(D,t)$
represent drift and volatility of $D_t$, respectively.  
In turn, the position of the particle, $\X_t$, obeys another
stochastic equation, 
\begin{equation}  \label{eq:dXt}
d\X_t = \sqrt{2D_t} \, d\bm{W}_t, 
\end{equation}
in which $\bm{W}_t = (W_t^1,\ldots,W_t^d)$ is formed by independent
Wiener processes $W_t^i$.  In this basic setting that we employ
throughout the paper, the particle undergoes locally isotropic
diffusion driven by instantaneous interactions with the thermal bath
(modeled by Gaussian noises $dW_t^i$) whose amplitude $\sqrt{2D_t}$
evolves with time.  To account for inert impermeable walls or
obstacles, a singular drift term should be added to the stochastic
equation (\ref{eq:dXt}), see \cite{Freidlin,Grebenkov06} for technical
details.  More generally, one can include local anisotropy and
external forces into Eq. (\ref{eq:dXt}) in a standard way
\cite{Risken}.  These extensions would change the governing
second-order differential operator and thus be incorporated via the
modified eigenvalues and eigenfunctions, as for homogeneous diffusion.
In contrast, the effect of rapid re-arrangements of the medium on
larger length scales is captured by the diffusing diffusivity $D_t$,
which is considered to be independent of local thermal noises
$dW_t^i$.  Most importantly, the stochastic equation (\ref{eq:dDt})
does not depend on the position $\X_t$ of the tracer.  As a
consequence, one can study separately the dynamics of the diffusivity
and then subordinate the dynamics of the tracer according to
Eqs. (\ref{eq:propagator}, \ref{eq:rho}).

For this purpose, one needs to evaluate the moment-generating function
$\Upsilon(t;\lambda|D_0,t_0)$, which is the Laplace transform of the
probability density $Q(t,T|D_0,t_0)$ of the integrated diffusivity
$T_t$ defined in Eq. (\ref{eq:Tt}):
\begin{eqnarray} \nonumber
\Upsilon(t;\lambda|D_0,t_0) &=& \E\biggl\{ \exp\biggl(-\lambda \int\limits_{t_0}^t dt' D_{t'}\biggr) \left|   D_{t_0} = D_0 \biggr\} \right. \\
&=& \int\limits_0^\infty dT \, e^{-\lambda T} \, Q(t;T|D_0,t_0) ,
\end{eqnarray}
given that the initial diffusivity at $t_0$ is $D_0$.  This function
satisfies the Feynman-Kac formula
\begin{equation}  \label{eq:FK}
\left(\partial_{t_0} + \mu(D_0,t_0) \partial_{D_0} 
+ \frac12 \sigma^2(D_0,t_0) \, \partial_{D_0}^2  - \lambda D_0 \right)  \Upsilon(t;\lambda|D_0,t_0) = 0 ,
\end{equation}
subject to the terminal condition 
\begin{equation}  \label{eq:terminal}
\Upsilon(t;\lambda|D_0,t_0 = t) = 1, 
\end{equation}
an appropriate boundary condition at $D_0 = 0$, and a regularity
condition $\Upsilon(t;\lambda|D_0,t_0) \to 0$ as $D_0 \to \infty$.
The boundary condition at $D_0 = 0$ should ensure that the diffusivity
remains nonnegative.  Following the discussion in
\cite{Lanoiselee18a,Lanoiselee18b}, we impose the no-flux boundary
condition to maintain the normalization of the probability density for
diffusivity.  For the backward equation, this condition reads
\begin{equation}  \label{eq:BC}
\left. \biggl(\partial_{D_0} \Upsilon(t;\lambda|D_0,t_0)\biggr)\right|_{D_0 = 0} = 0 .
\end{equation}

We note that the terminal condition (\ref{eq:terminal}) postulates the
average over the diffusivity at time $t$.  If one was interesting in
knowing the value of diffusivity at time $t$, $D_t = D$, the terminal
condition would be replaced by $\Upsilon(t;\lambda|D_0,t_0 = t) =
\delta(D - D_0)$.  In particular, substituting such
$\Upsilon(t;\lambda|D_0,t_0)$ into the spectral expansion
Eq. (\ref{eq:propagator}) would yield the full propagator
$P(\x,D,t|\x_0,D_0,t_0)$ characterizing both the position and the
diffusivity of the particle.  However, we do not consider this
extension in the paper.

In the remaining part of the paper, we focus on the case of diffusing
diffusivity that is homogeneous in time, i.e., the drift and
volatility coefficients are time-independent:
\begin{equation}
\mu(D,t) = \mu(D), \qquad \sigma(D,t) = \sigma(D).
\end{equation}
In this case, the solution of Eq. (\ref{eq:FK}) depends on the
difference $t - t_0$, i.e., $\Upsilon(t;\lambda|D_0,t_0) =
\Upsilon(t-t_0;\lambda|D_0,0)$, so that one can replace
$\partial_{t_0}$ by $- \partial_t$ and then set $t_0 = 0$:
\begin{equation}  \label{eq:FK2}
\left(\partial_{t} - \mu(D_0) \partial_{D_0} 
- \frac12 \sigma^2(D_0) \, \partial_{D_0}^2  + \lambda D_0 \right)  \Upsilon(t;\lambda|D_0) = 0 ,
\end{equation}
subject to the initial condition $\Upsilon(t=0;\lambda|D_0) = 1$ and
the same boundary condition (\ref{eq:BC}) and regularity condition
(note that we omitted $t_0 = 0$ in $\Upsilon(t;\lambda|D_0)$).  This
equation for the function $\Upsilon(t;\lambda|D_0)$ can also be used
to derive equations for the moments of the integrated diffusivity
$T_t$.  In a standard way, substituting the expansion
\begin{equation}  \label{eq:auxil11}
\Upsilon(t;\lambda|D_0) = \sum\limits_{n=0}^\infty \frac{(-1)^n}{n!} \lambda^n \langle T_t^n |D_0\rangle 
\end{equation}
into Eq. (\ref{eq:FK2}) and grouping the terms of the same order in
$\lambda$ yield a set of equations
\begin{equation}  \label{eq:Tmoments}
\fl
\left(\partial_t - \mu(D_0) \partial_{D_0} - \frac12 \sigma^2(D_0) \partial^2_{D_0} \right) \langle T_t^n |D_0\rangle 
= D_0 \langle T_t^{n-1} |D_0\rangle  \qquad (n = 1,2,\ldots),
\end{equation}
subject to the initial condition $\langle T_0^n |D_0\rangle = 0$ and
former boundary conditions.  In particular, the mean integrated
diffusivity obeys
\begin{equation}  \label{eq:Tmean}
\left(\partial_t - \mu(D_0) \partial_{D_0} - \frac12 \sigma^2(D_0) \partial^2_{D_0} \right) \langle T_t |D_0\rangle = D_0 .
\end{equation}

When there exists a unique equilibrium distribution of diffusivity,
$p_{\rm eq}(D)$, at which the probability flux of the associated
forward Fokker-Planck equation (with $\lambda = 0$) vanishes, i.e.
\begin{equation}  \label{eq:peq}
J_{\rm eq}(D) = \mu(D) p_{\rm eq}(D) - \partial_D \biggl( \frac{\sigma^2(D)}{2} \, p_{\rm eq}(D) \biggr) = 0 ,
\end{equation}
it is convenient to average over the initial diffusivity $D_0$ drawn
from this equilibrium density:
\begin{equation}
\Upsilon(t;\lambda) = \int\limits_0^\infty dD_0 \, \Upsilon(t;\lambda|D_0) \, p_{\rm eq}(D_0).
\end{equation}

\subsection{Example: a Feller process}
\label{sec:Feller}

In \cite{Lanoiselee18a,Lanoiselee18b}, we studied in detail the
diffusing diffusivity modeled by a Feller process, for which
\begin{equation}  \label{eq:Feller}
\mu(D,t) = (\bar{D} - D)/\tau , \qquad \sigma(D,t) = \sigma \sqrt{2D} ,
\end{equation}
with three parameters: the mean diffusivity $\bar{D}$, the relaxation
time scale $\tau$, and the amplitude of diffusivity fluctuations
$\sigma$.  In particular, we obtained%
\footnote{
Three misprints were found in \cite{Lanoiselee18a} in inline equations
after Eq. (6), compare them with our corrected
Eq. (\ref{eq:Upsilon_D0}).  These misprints did not affect the
remaining content of Ref. \cite{Lanoiselee18a}. }
%
\begin{eqnarray}    \label{eq:Upsilon_D0}
&& \Upsilon(t; \lambda|D_0) = \left(\frac{2\omega e^{-(\omega-1)t/(2\tau)}}{\omega+1 + (\omega-1)e^{-\omega t/\tau}}\right)^\nu \\
\nonumber
&& \times \exp\left(\frac{D_0 (\omega+1)}{2\sigma^2\tau} \left(1 - \frac{2\omega}{\omega+1 + (\omega-1)e^{-\omega t/\tau}}\right)\right)
\end{eqnarray}
and
\begin{equation}  \label{eq:Ups}
\Upsilon(t; \lambda) = \left( \frac{4\omega e^{-(\omega-1) t/(2\tau)}}{(\omega+1)^2 - (\omega-1)^2 e^{-\omega t/\tau} } \right)^\nu ,
\end{equation}
with $\omega = \sqrt{1 + 4\sigma^2\tau^2 \lambda}$, $\nu =
\bar{D}/(\tau \sigma^2)$, and the equilibrium diffusivity 
is known to follow the Gamma distribution:
\begin{equation}  \label{eq:Gamma}
p_{\rm eq}(D) = \frac{\nu^\nu D^{\nu-1}}{\Gamma(\nu) \bar{D}^\nu} \exp(-\nu D/\bar{D}).
\end{equation}
The first-passage and extreme value properties of the Feller process
itself were studied earlier in \cite{Masoliver12,Masoliver14,Gan15}.

\subsection{Example: reflected Brownian motion}
\label{sec:rBm}

Chubynsky and Slater first introduced the diffusing diffusivity
qualitatively as reflected Brownian motion on the positive half-line
\cite{Chubynsky14}.  To avoid an unlimited growth of diffusivity, it
is more natural to consider Brownian motion on an interval $(0,D_m)$
with two reflecting endpoints.  We explore this case with
\begin{equation}
\mu(D,t) = 0, \qquad \sigma(D,t) = \sigma ,
\end{equation}
so that Eq. (\ref{eq:FK2}) becomes
\begin{equation}  \label{eq:FK_CS}
\biggl(\partial_{t} - \frac{\sigma^2}{2} \, \partial_{D_0}^2  + \lambda D_0 \biggr)  \Upsilon(t;\lambda|D_0) = 0 ,
\end{equation}
subject to the initial condition $\Upsilon(t=0;\lambda|D_0) = 1$ and
two boundary conditions:
\begin{equation}  \label{eq:BC_uniform}
\left. \biggl(\partial_{D_0} \Upsilon(t;\lambda|D_0) \biggr) \right|_{D_0 = 0} =
\left. \biggl(\partial_{D_0} \Upsilon(t;\lambda|D_0) \biggr) \right|_{D_0 = D_m} = 0 .
\end{equation}
The amplitude of fluctuations, $\sigma$, strongly affects the
diffusivity dynamics.  In the limit $\sigma \to 0$, fluctuations are
suppressed, and one deals with a constant initial diffusivity $D_0$.
In the opposite limit $\sigma\to\infty$, the diffusivity switches so
rapidly between different values in $(0,D_m)$ that its behavior
resembles a constant mean diffusivity $D_m/2$.

The equation (\ref{eq:FK_CS}) admits a standard spectral solution in
terms of eigenvalues and eigenfunctions of the associated differential
operator $\L = \partial_{D_0}^2 - \alpha^3 D_0$ (with $\alpha^3 =
2\lambda/\sigma^2$).  One can search for such an eigenpair $\gamma$
and $v(D_0)$ in the form
\begin{equation}  \label{eq:vn_uniform}
v(D_0) = C \biggl[\Bi'(-\beta \gamma) \, \Ai\bigl(\alpha D_0 - \beta \gamma\bigr) 
- \Ai'( -\beta \gamma)\, \Bi\bigl(\alpha D_0 - \beta \gamma\bigr)\biggr],
\end{equation}
where $\Ai(z)$ and $\Bi(z)$ are two linearly independent Airy
functions, prime denotes the derivative, $\beta = 1/\alpha^2$, and $C$
is a normalization constant which is fixed by imposing the
$L_2$-normalization of the eigenfunction:
\begin{equation}
\int\limits_{0}^{D_m} dD \, v^2(D) = 1 
\end{equation}
(see Refs. \cite{Stoller91,Grebenkov14} for a more detailed analysis
of a similar problem).  This integral can be evaluated using the Airy
equation and the reflecting boundary conditions:
\begin{equation}
\int\limits_{0}^{D_m} dD \, v^2(D) =  \frac{(\alpha D_m - \beta \gamma) v^2(D_m) + \beta \gamma v^2(0)}{\alpha} \,,
\end{equation}
from which $C$ can be expressed as
\begin{equation}
C = \pi \sqrt{\alpha} \left(\beta \gamma + (\alpha D_m - \beta \gamma) 
\left(\frac{\Bi'(-\beta \gamma)}{\Bi'(\alpha D_m - \beta \gamma)}\right)^2 \right)^{-1/2} \,,
\end{equation}
where we used the Wronskian of Airy functions and reflected boundary
conditions (note also that $v(0) = C/\pi$ and $v(D_m) = C \Bi'(-\beta
\gamma)/(\pi \Bi'(\alpha D_m - \beta \gamma))$).

The form (\ref{eq:vn_uniform}) already satisfies the reflecting
boundary condition at $D_0 = 0$.  The eigenvalue $\gamma$ is
determined from the second boundary condition at $D_0 = D_m$ that
implies
\begin{equation}  \label{eq:eigen_gamma}
\Ai'\bigl( - \beta \gamma\bigr) \Bi'\bigl(\alpha D_m - \beta \gamma\bigr) 
- \Ai'\bigl(\alpha D_m - \beta \gamma\bigr) \Bi'\bigl( - \beta \gamma\bigr) = 0.
\end{equation}
As $-\alpha^3 D_0$ is a bounded perturbation of the double derivative
operator on an interval, the spectrum of the operator $\L$ is
discrete, i.e., there are infinitely many solutions $\gamma_k$ of
Eq. (\ref{eq:eigen_gamma}) that we enumerate by index $k =
0,1,2,\ldots$.  The associated eigenfunctions $v_k$ form a complete
orthonormal basis in $L_2(0,D_m)$.  As a consequence, the solution of
Eq. (\ref{eq:FK_CS}) can be decomposed on this basis as
\begin{equation}  \label{eq:Ups_uniform_D0}
\Upsilon(t;\lambda|D_0) = \sum\limits_{k=0}^\infty e^{- \sigma^2 t \gamma_k/2} v_k(D_0)  \int\limits_{0}^{D_m} dD \, v_k(D).
\end{equation}
If the initial diffusivity $D_0$ is drawn from the equilibrium density
$p_{\rm eq}$ (which is uniform in this setting), one gets
\begin{equation}  \label{eq:Ups_uniform}
\Upsilon(t;\lambda) = \frac{1}{D_m} \sum\limits_{k=0}^\infty e^{- \sigma^2 t \gamma_k/2} \left(\int\limits_{0}^{D_m} dD \, v_k(D)\right)^2 .
\end{equation}
From the moment-generating function, one can compute the moments of
the integrated diffusivity.  One can either perform the asymptotic
analysis of Eq. (\ref{eq:Ups_uniform_D0}) as $\lambda \to 0$, or solve
directly Eq. (\ref{eq:Tmean}) with $\mu = 0$, subject to the
reflecting boundary conditions at $0$ and $D_m$.  In the latter case,
an expansion of $\langle T_t | D_0\rangle$ over the complete basis of
cosine functions on $(0,D_m)$ leads to
\begin{equation} 
\fl
\langle T_t | D_0\rangle = \frac{D_m}{2} t + \frac{4D_m^3}{\sigma^2} \sum\limits_{n=1}^\infty 
\frac{(1-(-1)^n)(1 - e^{-\pi^2 n^2 \sigma^2 t/(2D_m^2)})}{\pi^4 n^4} \cos(\pi n D_0/D_m) \,.
\end{equation}
In the short-time limit, one retrieves $\langle T_t | D_0\rangle
\simeq D_0 t$, whereas in the long-time limit, one gets
\begin{equation}
\langle T_t | D_0\rangle \simeq \frac{D_m}{2} t + \frac{4D_0^3 - 6 D_0^2 D_m + D_m^3}{12 \sigma^2} \,,
\end{equation}
with the expected dominant behavior $D_m t/2$.  Higher-order moments
obeying Eqs. (\ref{eq:Tmoments}) can be found in the same way.

In sharp contrast to the fully explicit solution (\ref{eq:Upsilon_D0})
for the Feller process, the solution (\ref{eq:Ups_uniform_D0})
requires a numerical computation of eigenvalues $\gamma_k$ which
depend implicitly on the parameter $\lambda$.  As the propagator
$P(\x,t|\x_0)$ in Eq. (\ref{eq:propagator}) also involves a spectral
decomposition over the Laplacian eigenvalues $\lambda_n$ in a bounded
domain, the eigenvalues $\gamma_k$ should be evaluated for each
$\lambda_n$ that makes this solution computationally demanding and
impractical.  At the same time, this solution allows one to analyze
the asymptotic behavior of $\Upsilon(t;\lambda)$ and all related
quantities as we briefly illustrate below.

For small $\lambda$, the term $-\alpha^3 D_0$ can be considered as a
small perturbation of the double derivative in the operator $\L =
\partial^2_{D_0} - \alpha^3 D_0$ so that the eigenvalues and
eigenfunctions of $\L$ are close to that of the double derivative
operator.  The perturbation theory yields thus
\begin{eqnarray}  \nonumber
\gamma_k &=& \pi^2 k^2/D_m^2 + \frac{2 - \delta_{k,0}}{D_m} \int\limits_0^{D_m} dD\, \cos^2(\pi k D/D_m) \, \alpha^3 D + O(\lambda^2) \\
\label{eq:gamma_small}
& =& \pi^2 k^2/D_m^2 + \lambda D_m/\sigma^2 + O(\lambda^2)  \qquad (\lambda \ll \sigma^2/D_m^3).
\end{eqnarray}
As expected, the correction terms are small for the modes with $k =
1,2,\ldots$, whereas the correction term is dominant for the constant
mode with $k = 0$.  In this limit, one gets
\begin{equation}
\Upsilon(t;\lambda) \simeq e^{- \sigma^2 t \gamma_0/2} + O(\lambda) = e^{-\lambda t D_m/2} + O(\lambda),
\end{equation}
given that the contribution of other terms is small because the
(unperturbed) eigenfunctions $\cos(\pi k D/D_m)$ are orthogonal to
$1$.  This analysis is also applicable in the limit $\sigma\to\infty$,
in which fluctuations are so strong that the model is reduced to
homogeneous diffusion with the mean diffusivity $D_m/2$.

In the opposite limit of large $\lambda$, one deals with large
$\alpha$ and small $\beta$ in Eq. (\ref{eq:eigen_gamma}) so that the
function $\Bi'\bigl(\alpha D_m - \beta \gamma\bigr)$ is exponentially
large, whereas the function $\Ai'\bigl(\alpha D_m - \beta
\gamma\bigr)$ is exponentially small.  As a consequence, zeros of
Eq. (\ref{eq:eigen_gamma}) are very close to the zeros of $\Ai'\bigl(-
\beta \gamma\bigr)$, i.e.,
\begin{equation}  \label{eq:gamma_large}
\gamma_k \simeq |a'_k| (\sigma^2/2)^{-2/3} \lambda^{2/3}  \qquad (\lambda \gg \sigma^2/D_m^3),
\end{equation}
where $a'_k$ are the zeros of the derivative of the Airy function
(e.g., $a'_0 \approx - 1.0188$).  We conclude that both
$\Upsilon(t;\lambda|D_0)$ and $\Upsilon(t;\lambda)$ decay with
$\lambda$ in a stretched-exponential way.

One can see that the ratio $\sigma^2/D_m^3$, setting the borderline
between two asymptotic limits (\ref{eq:gamma_small},
\ref{eq:gamma_large}), introduces a characteristic length of dynamic
disorder, $\sqrt{D_m^3/\sigma^2}$.  This length scale is compared in
Eqs. (\ref{eq:propagator}, \ref{eq:rho}) to the diffusion length
$\sqrt{D_m t}$ and to the geometric length scales of the reactive
medium determined by the eigenvalues $\lambda_n^{-1/2}$.  In
particular, various asymptotic limits of the first-passage time
density can be deduced, in analogy with the results presented in
\cite{Lanoiselee18b} for the case of the diffusivity modeled by a
Feller process.

It is instructive to look at the limit $\sigma = 0$, in which the
diffusivity does not fluctuate, so that $\Upsilon(t;\lambda|D_0) =
e^{-D_0 \lambda t}$, where $D_0$ is the initial diffusivity.  If this
diffusivity is randomly chosen from the equilibrium distribution, one
gets 
\begin{equation}  \label{eq:Ups_sigma0}
\Upsilon(t;\lambda) = \frac{1 - e^{-D_m t\lambda}}{D_m t\lambda} \,.
\end{equation}
As a consequence, when either $t$ or $\lambda$ goes to infinity, the
function $\Upsilon(t;\lambda)$ decays slowly (as a power law).  This
slow decay is a consequence of the superstatistical description: the
average over $D_0$ includes the contribution from particles with
arbitrarily small diffusivities.  This is drastically different from
the stretchted-exponential decay with respect to $\lambda$ and from the
exponential decay with respect to $t$ in the presence of fluctuations:
even though small diffusivities are still accessible, it is unlikely
that the particle keeps such a small diffusivity for a long time.

We also note that the solution of a more general problem of reflected
Brownian motion on a shifted interval $(D_1,D_1+D_m)$ (with $D_1 > 0$)
can be easily reduced to our solution by shifting the diffusivity
$D_t$, i.e., by considering $\hat{D}_t = D_t + D_1$, where $D_t$ is
modeled by reflected Brownian motion on $(0,D_m)$ as before.  The
shift by $D_1$ leads to a constant term $D_1 t$ in the integrated
diffusivity so that $\hat{\Upsilon}(t;\lambda) = e^{-D_1 t \lambda}
\Upsilon(t;\lambda)$.  We note that the explicit factor $e^{-D_1 t
\lambda}$ provides the dominant contribution to the decrease of the 
function $\hat{\Upsilon}(t;\lambda)$ as compared to
$\Upsilon(t;\lambda)$ in the limit $\lambda\to\infty$.

\subsection{Moments of the position}

In \cite{Lanoiselee18a}, the propagator $P(x,t|x_0)$ for diffusion on
the line (without boundary) was expressed in terms of the
moment-generating function $\Upsilon(t;\lambda)$ for the Feller
process:
\begin{equation}  \label{eq:prop_Ups}
P(x,t|x_0) = \int\limits_{-\infty}^\infty \frac{dq}{2\pi} \, e^{iq(x-x_0)} \, \Upsilon(t;q^2) .
\end{equation}
The subordination argument \cite{Lanoiselee18b,Chechkin17} supports
this relation for any model of diffusing diffusivity.  This relation
provides thus an additional interpretation of $\Upsilon(t;\lambda)$ as
the characteristic function of the one-dimensional displacement on a
line.  In particular, one can easily evaluate the moments of the
displacement, e.g.,
\begin{eqnarray}
\langle X_t \rangle &=& -i \biggl(\partial_q \Upsilon(t;q^2) \biggr)_{q=0} = -i
\biggl(2\sqrt{\lambda}\, \partial_\lambda \Upsilon(t;\lambda) \biggr)_{\lambda =0} = 0 ,\\
\langle X_t^2 \rangle &=& - \biggl(\partial^2_q \Upsilon(t;q^2) \biggr)_{q=0} 
= -2 \biggl(\partial_\lambda \Upsilon(t;\lambda)\biggr)_{\lambda =0} ,\\
\langle X_t^3 \rangle &=& i \biggl(\partial^3_q \Upsilon(t;q^2) \biggr)_{q=0} 
= i \biggl(12 \sqrt{\lambda} \partial^2_\lambda \Upsilon(t;\lambda) + 8\lambda^{\frac32} \partial^3_\lambda \Upsilon(t;\lambda) \biggr)_{\lambda =0} = 0 ,\\
\langle X_t^4 \rangle &=& \biggl(\partial^4_q \Upsilon(t;q^2) \biggr)_{q=0} 
= 12 \biggl( \partial^2_\lambda \Upsilon(t;\lambda)  \biggr)_{\lambda =0} .
\end{eqnarray}
As expected, odd moments vanish due to the symmetry of thermal noise
$dW_t$ (and independently of the diffusivity model), whereas even
moments can be expressed through the moments of the integrated
diffusivity
\begin{equation}
\langle X_t^{2n} \rangle = \frac{(2n)!}{n!}\, \langle T_t^n \rangle .
\end{equation}
This is an extension of the basic relation for the moments of the
homogeneous Gaussian diffusion, for which $\langle T_t^n \rangle =
(D_0 t)^n$.  From this relation, one easily gets the kurtosis, as well
as the non-Gaussian parameter, $\gamma(t) = \langle
X_t^4\rangle/(3\langle X_t^2\rangle^2) - 1$.  Note that the same
relation holds for the moments with a prescribed initial diffusivity
$D_0$ which can be found by solving Eqs. (\ref{eq:Tmoments}):
\begin{equation}
\langle X_t^{2n} | D_0 \rangle = \frac{(2n)!}{n!}\, \langle T_t^n | D_0\rangle .
\end{equation}
In contrast, there is no such a direct relation between the moments
$\langle T_t^n \rangle$ and $\langle X_t^{2n} \rangle$ for restricted
diffusion.

\section{Switching diffusion model}
\label{sec:switching}

A switching diffusion model, in which a diffusing particle randomly
switches between internal states with distinct diffusivities, can be
considered as a discrete version of diffusing diffusivity models.
According to the subordination argument \cite{Lanoiselee18b}, it is
enough to obtain the propagator for one-dimensional switching
diffusion on a line, see Eq. (\ref{eq:prop_Ups}), whereas its Fourier
transform yields the function $\Upsilon(t;\lambda)$ and thus accesses
general FPT problems in arbitrary confined reactive domains in $\R^d$
via Eqs. (\ref{eq:propagator}, \ref{eq:rho}).  We consider such a
model with $J$ states which are characterized by a set of diffusion
coefficients $D_i$ and switching rates $k_{ij}$ (a rigorous
mathematical formulation of switching models and some their properties
can be found in \cite{Bressloff17,Yin10,Yin10b,Baran13}).  We
introduce the probability density $P_{i,i_0}(x,t|x_0,t_0)$ of finding
the particle in a vicinity of $x$ in the state $i$ at time $t$, given
that it was started from $x_0$ in the state $i_0$ at time $t_0$.  This
propagator satisfies the forward Fokker-Planck equation
\begin{equation}  \label{eq:prop_auxil}
\partial_t P_{i,i_0} = D_i \partial_x^2 P_{i,i_0} + \sum\limits_{j=1}^J k_{ji} P_{j,i_0} ,
\end{equation}
subject to the initial condition: $P_{i,i_0}(x,t=t_0|x_0,t_0) =
\delta_{i,i_0} \delta(x-x_0)$, where we defined $k_{ii} = - \sum\limits_{j\ne i}
k_{ij}$.  The first term on the right-hand side describes diffusion
(with diffusivity $D_i$), while the second term accounts for switching
between different states.  In this class of switching models, the
dynamics in each internal state is governed by the same differential
operator and differs only by its diffusivity.  This is the crucial
property that will allow for getting the propagator $P(\x,t|\x_0)$ in
Eq. (\ref{eq:propagator}) for this model.  Such an extension is not
directly applicable to other intermittent processes, in which the
governing operator changes between states.  Moreover, modifications
are needed even in the case when the operator remains the same (e.g.,
the Laplace operator) but the boundary condition changes between
states (see, e.g., a two-state model developed in \cite{Godec17}, in
which the particle reacts with the target only when it is in an
``active'' state).  Similarly, the models of surface-mediated
diffusion
\cite{Benichou10a,Benichou11a,Rojo11,Rupprecht12a,Rupprecht12b} are
not considered here as their switching mechanisms are different.

The Fourier transform reduces the partial differential equations
(\ref{eq:prop_auxil}) to a set of linear ordinary differential
equations that can be solved in a matrix form, from which
\begin{equation}
P_{i,i_0}(x,t|x_0,t_0) = \int\limits_{-\infty}^\infty \frac{dq}{2\pi}\, e^{-iq(x-x_0)}  
\biggl[\exp\bigl(-(q^2\D - \K^\dagger)(t-t_0)\bigr)\biggr]_{i,i_0} ,
\end{equation}
where $\D$ is the diagonal $J\times J$ matrix of diffusivities,
$\D_{ij} = \delta_{ij} D_i$, and $\K$ is the matrix of switching
rates, $\K_{ij} = k_{ij}$.  If $p_i$ denotes the probability of
starting in the initial state $i$, the marginal propagator averaged
over the initial and arrival states reads
\begin{equation}
P(x,t|x_0,t_0) = \int\limits_{-\infty}^\infty \frac{dq}{2\pi}\, e^{-iq(x-x_0)} \, \Upsilon(t-t_0;q^2),
\end{equation}
where
\begin{equation}  \label{eq:Y_switching}
\Upsilon(t;\lambda) = \left(\begin{array}{c} 1 \\ 1 \\ ... \\ 1 \\ \end{array} \right)^\dagger 
\exp\bigl(-(\lambda \D - \K^\dagger)t\bigr)  \left(\begin{array}{c} p_1 \\ p_2 \\ ... \\ p_J \\ \end{array} \right).
\end{equation}
Note that the function $\Upsilon(t;\lambda|D_0)$ admits the same form,
with $p_j$ being equal to $0$ for all states, except for the state
with $D_0$ (for which $p_j = 1$). 

Using this relation, one can access the propagator $P(\x,t|\x_0)$ and
the first-passage time density $\rho(t|x_0)$ in a general confining
domain according to Eqs. (\ref{eq:propagator}, \ref{eq:rho}).  We
recall that the matrix exponential function in
Eq. (\ref{eq:Y_switching}) can be evaluated via diagonalization of the
matrix $\lambda \D - \K^\dagger$.  A fully explicit solution can be
obtained for the two-state switching model (see, e.g.,
\cite{Lanoiselee18a,Karger85}):
\begin{equation}
\Upsilon(t;\lambda) = \frac{e^{-\gamma_+ t}(\bar{D} \lambda - \gamma_-) - e^{-\gamma_- t}(\bar{D} \lambda - \gamma_+)}{\gamma_+ - \gamma_-} \,,
\end{equation}
where $\bar{D} = p_1 D_1 + p_2 D_2$ and
\begin{eqnarray}  
\gamma_\pm &=& \frac12 \biggl((D_1+D_2)\lambda + (k_{12} + k_{21}) \\   \nonumber
&\pm& \sqrt{((D_2-D_1)\lambda + (k_{21}-k_{12}))^2 + 4k_{12} k_{21}} \biggr) .
\end{eqnarray}
This rigorous result refines a former discussion of the two-state
noise in \cite{Tyagi17}.  For a larger number of states, formulas
rapidly become too cumbersome and impractical.  In contrast, a
numerical computation of the function $\Upsilon(t;\lambda)$ via the
matrix form (\ref{eq:Y_switching}) remains efficient even for
relatively large number of states (up to few thousand).

\subsection{Relation to diffusing diffusivity models}

In this subsection, we discuss how discretized versions of diffusing
diffusivity models with time-independent coefficients $\mu(D)$ and
$\sigma(D)$ are related to switching diffusion models.  In fact,
continuously varying diffusivity $D_t$ can be replaced by a set of
discrete values $D_i = i \, \dD$ ($i = 1,2,\ldots$), with a
discretization step $\dD$.  Variations of $D_t$ can thus be seen as
switching between neighboring states.  We briefly discuss two
equivalent approaches to formalize this connection.

In the first approach, the switching rates are determined by
discretizing the forward Fokker-Planck equation for the probability
density $p(D,t|D_0,t_0)$ of the diffusivity $D_t$ (in the It\^o
convention)
\begin{equation}
\partial_t p(D,t|D_0,t_0) = - \partial_D \bigl(\mu(D) p\bigr) + \frac12 \partial_D^2 \bigl(\sigma^2(D) p\bigr),
\end{equation}
subject to the initial condition $p(D,t=t_0|D_0,t_0) = \delta(D-D_0)$.
The discretization of the right-hand side of this equation with a
diffusivity step $\dD$ reads
\begin{eqnarray} \nonumber
 \partial_t p(D,t) &=& \frac{\sigma^2(D+\dD)}{2\dD^2} p(D+\dD,t) 
- \left[\frac{\sigma^2(D)}{\dD^2} + \frac{\mu(D)}{\dD}\right] p(D,t) \\  \nonumber
&+& \left[\frac{\sigma^2(D-\dD)}{2\dD^2} + \frac{\mu(D-\dD)}{\dD}\right] p(D-\dD,t).
\end{eqnarray}
In other words, the original PDE is approximated by a set of linear
ordinary differential equations.  These discretized equations for
$p(i \dD,t|D_0,t_0)$ can be seen as a switching model with
multiple states $D_i = i\dD$ ($i = 1,2,\ldots$) and the switching
rates
\begin{equation}  \label{eq:rates_auxil}
k_{i+1,i} = \frac{\sigma^2_{i+1}}{2\dD^2}  \,, \qquad 
k_{i-1,i} = \frac{\sigma^2_{i-1}}{2\dD^2} + \frac{\mu_{i-1}}{\dD}  \,, \qquad
k_{i,i} =  - \frac{\sigma^2_i}{\dD^2} - \frac{\mu_i}{\dD}  \,,  
\end{equation}
and zero otherwise, where we used the shortcut notations $\sigma_i =
\sigma(i\dD)$ and $\mu_i = \mu(i\dD)$.  As the ``first'' equation for
$p(\dD,t)$ involves the term $p(0,t)$, one needs to close this system
by accounting for the reflecting boundary condition at $D = 0$:
\begin{equation}
\fl
0 = J(0) = \biggl(\mu p - \frac12 \partial_D (\sigma^2 p)\biggr)\biggl|_{D=0} = \mu(0) p(0,t) 
- \frac{\sigma^2(\dD) p(\dD,t) - \sigma^2(0) p(0,t)}{2\dD} \,.
\end{equation}
Expressing $p(0,t)$ in terms of $p(\dD,t)$ leads to a slight
modification of the first diagonal element: $k_{1,1} = -
\sigma^2_1/(2\dD^2) - \mu_1/\dD$.  

As a numerical solution of an infinitely-dimensional system of
equations is not feasible, one needs to truncate the original problem
by imposing an additional reflecting boundary condition at some
truncation level $D_m$.  As in the case of $D = 0$, this boundary
condition changes the coefficient $k_{J,J}$ in the ``last'' equation
for $p(D_m,t)$, with $D_m = J \dD$.  However, a simpler and more
consistent way of closing the system of equations is to require
$\sum\nolimits_j k_{ij} = 0$ for $i = J$.  This is a detailed balance
condition for the matrix of switching rates, which is already
satisfied for all rates from Eq. (\ref{eq:rates_auxil}) with
$i=1,\ldots,J-1$.
Combining these relations, one can write the set of discretized
equations in a matrix form as 
\begin{equation}
\partial_t \left(\begin{array}{c} p(\dD,t) \\ p(2\dD,t) \\ ... \\ p(J\dD,t) \end{array}\right) 
= \K^\dagger \left(\begin{array}{c} p(\dD,t) \\ p(2\dD,t) \\ ... \\ p(J\dD,t) \end{array}\right) ,
\end{equation}
with the $J \times J$ three-diagonal matrix
\begin{equation}
\fl
\K = \left(\begin{array}{c c c c c c} 
- \frac{\sigma_1^2}{2\dD^2} - \frac{\mu_1}{\dD} & \frac{\sigma_1^2}{2\dD^2} + \frac{\mu_1}{\dD} & 0 & ... & 0 & 0 \\
\frac{\sigma_2^2}{2\dD^2} & - \frac{\sigma_2^2}{\dD^2} - \frac{\mu_2}{\dD} 
& \frac{\sigma_2^2}{2\dD^2} + \frac{\mu_2}{\dD} & ... & 0 & 0 \\
0 &   \frac{\sigma_3^2}{2\dD^2} & -\frac{\sigma_3^2}{\dD^2} - \frac{\mu_3}{\dD} & ... & 0 & 0 \\
... & ... & ... & ... & ... & ... \\
 0 & 0  & 0 & ... & -\frac{\sigma_{J-1}^2}{\dD^2} - \frac{\mu_{J-1}}{\dD} & \frac{\sigma_{J-1}^2}{2\dD^2} + \frac{\mu_{J-1}}{\dD} \\
 0 & 0  & 0 & ... & \frac{\sigma_J^2}{2\dD^2} & -\frac{\sigma_J^2}{2\dD^2} \\  \end{array} \right),
\end{equation}
whereas $\D_{ij} = \delta_{ij} i \dD$.  In other words, we identified
the matrices $\D$ and $\K$ determining a switching diffusion model
that is a discrete approximation of the diffusing diffusivity model
with coefficients $\mu(D)$ and $\sigma(D)$.  The relation
(\ref{eq:Y_switching}) expresses the function $\Upsilon(t;\lambda)$
for this switching model.

Alternatively, one could directly discretize the backward equation
(\ref{eq:FK}):
\begin{eqnarray} \nonumber
&& - \partial_{t_0} \Upsilon(t;\lambda|D_0,t_0) = \left[\frac{\sigma^2(D_0)}{2\dD^2} 
+ \frac{\mu(D_0)}{\dD}\right]  \Upsilon(t;\lambda|D_0+\dD,t_0)  \\  \nonumber
&& - \left[\frac{\sigma^2(D_0)}{\dD^2} + \frac{\mu(D_0)}{\dD} + \lambda D_0\right] \Upsilon(t;\lambda|D_0,t_0) 
+ \frac{\sigma^2(D_0)}{2\dD^2} \Upsilon(t;\lambda|D_0-\dD,t_0)
\end{eqnarray}
(here for convenience we adopted another discretization scheme for the
first derivative).  The solution of this discretized equation with the
terminal condition $\Upsilon(t;\lambda|D_0,t_0=t) = 1$ reads
\begin{equation}
\Upsilon(t;\lambda|i \dD,t_0) = \left[\exp(- (\lambda \D - \K)(t-t_0))  \left(\begin{array}{c} 1 \\ 1 \\ ... \\ 1 \\ \end{array} \right)\right]_i ,
\end{equation}
with the matrices $\D$ and $\K$ defined above.  Averaging this
solution over the initial states chosen with probabilities $p_j$ and
setting $t_0 = 0$ yield
\begin{equation}
\Upsilon(t;\lambda) = \left(\begin{array}{c} p_1 \\ p_2 \\ ... \\ p_J \\ \end{array} \right)^\dagger \exp(- (\lambda \D - \K)t) 
 \left(\begin{array}{c} 1 \\ 1 \\ ... \\ 1 \\ \end{array} \right) ,
\end{equation}
which is just a transposed re-writing of Eq. (\ref{eq:Y_switching}).

\subsection{Numerical illustrations}

\begin{figure}
\begin{center}
\includegraphics[width=100mm]{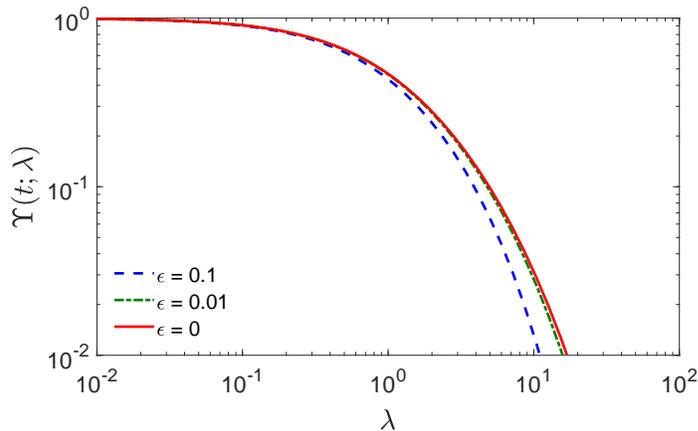}
\end{center}
\caption{
Comparison of the function $\Upsilon(t;\lambda)$ for the Feller
diffusing diffusivity model and its approximations by switching
diffusion models.  Drift and volatility coefficients are set by
Eq. (\ref{eq:Feller}) with parameters $\bar{D} = 1$, $\sigma = 1$,
$\tau = 1$, and $t = 1$ (arbitrary units).  Solid line shows the
explicit solution (\ref{eq:Ups}) whereas two other lines present
approximate solutions from Eq. (\ref{eq:Y_switching}) for switching
diffusion models with $\dD = 0.1$ (dashed line) and $\dD = 0.01$
(dash-dotted line) and $D_m = 10$.  }
\label{fig:Upsilon}
\end{figure}

Figure \ref{fig:Upsilon} illustrates a comparison between a diffusing
diffusivity model and its approximation by switching diffusion.  We
set the drift and the volatility terms according to
Eq. (\ref{eq:Feller}) for the Feller process, with $\bar{D} = 1$,
$\sigma = 1$, and $\tau = 1$ (arbitrary units).  On one hand, the
function $\Upsilon(t;\lambda)$ is computed via explicit analytical
solution (\ref{eq:Ups}).  On the other hand, a discrete approximation
of this process by switching diffusion allows one to compute the
function $\Upsilon(t;\lambda)$ by Eq. (\ref{eq:Y_switching}).  This
computation depends on the discretization step $\dD$ and the
truncation threshold $D_m$.  We set $D_m = 10$ and checked that
further increase of this value does not almost affect the computation.
This is not surprising given that the equilibrium distribution of
diffusivities, Eq. (\ref{eq:Gamma}), decays exponentially fast for the
Feller process.  The discretization step has also relatively weak
impact on the solution, if it is small enough (see
Fig. \ref{fig:Upsilon}).  We emphasize, however, the quality of the
approximation depends in general on the chosen model and its
parameters.  For instance, setting $\tau = 10$ (while keeping $\bar{D}
= 1$ and $\sigma = 1$) yields $\nu = 0.1$ in the Feller model and thus
an integrable but divergent at $D = 0$ equilibrium density in
Eq. (\ref{eq:Gamma}).  As a consequence, a much finer discretization
is needed to accurately capture the behavior of this density near zero
and thus to get an accurate representation via a switching diffusion
model.  In general, one needs to undertake the convergence analysis or
at least to compute $\Upsilon(t;\lambda)$ with various discretization
steps $\dD$ to check its convergence.

\begin{figure}
\begin{center}
\includegraphics[width=70mm]{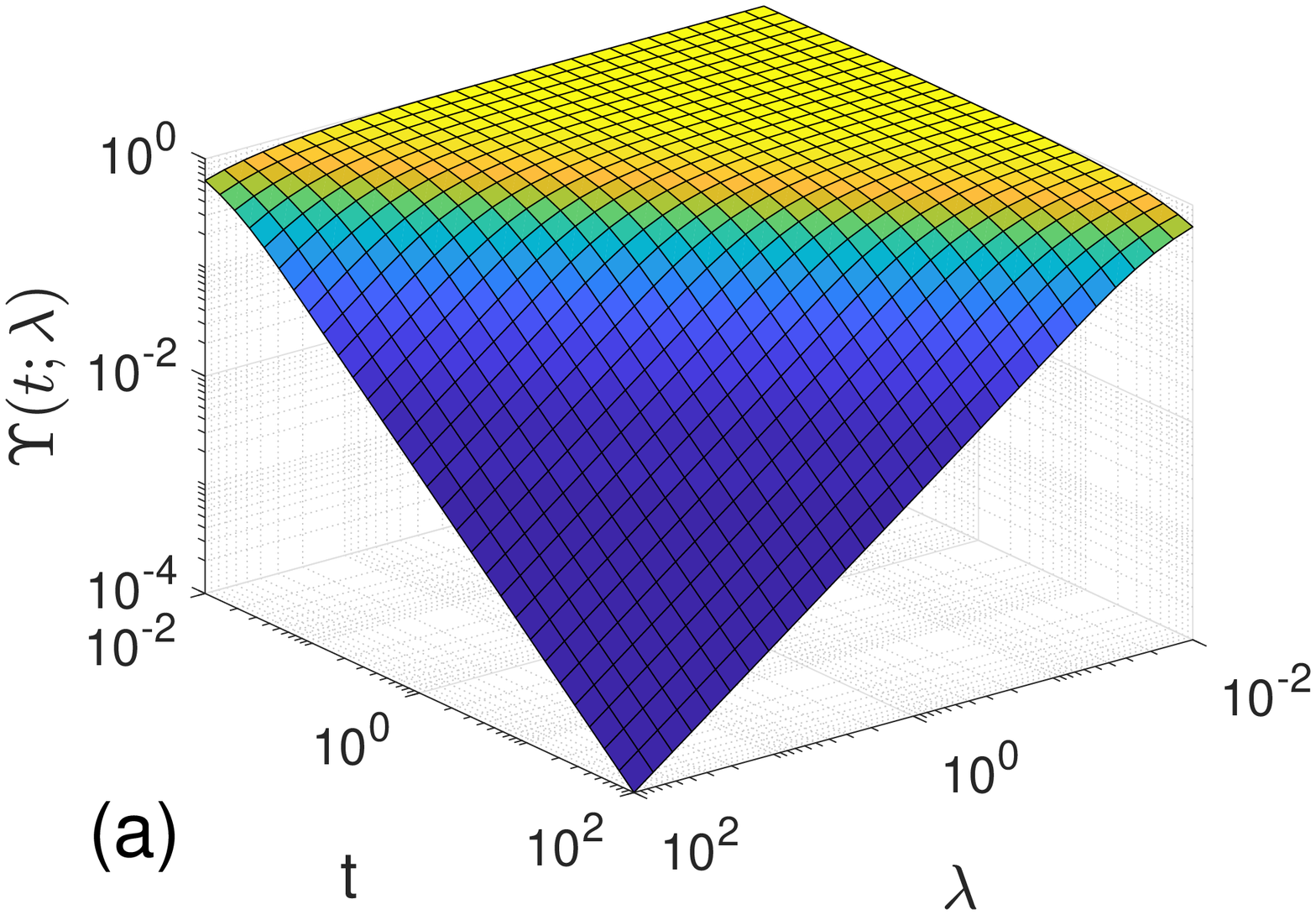}
	\includegraphics[width=82mm]{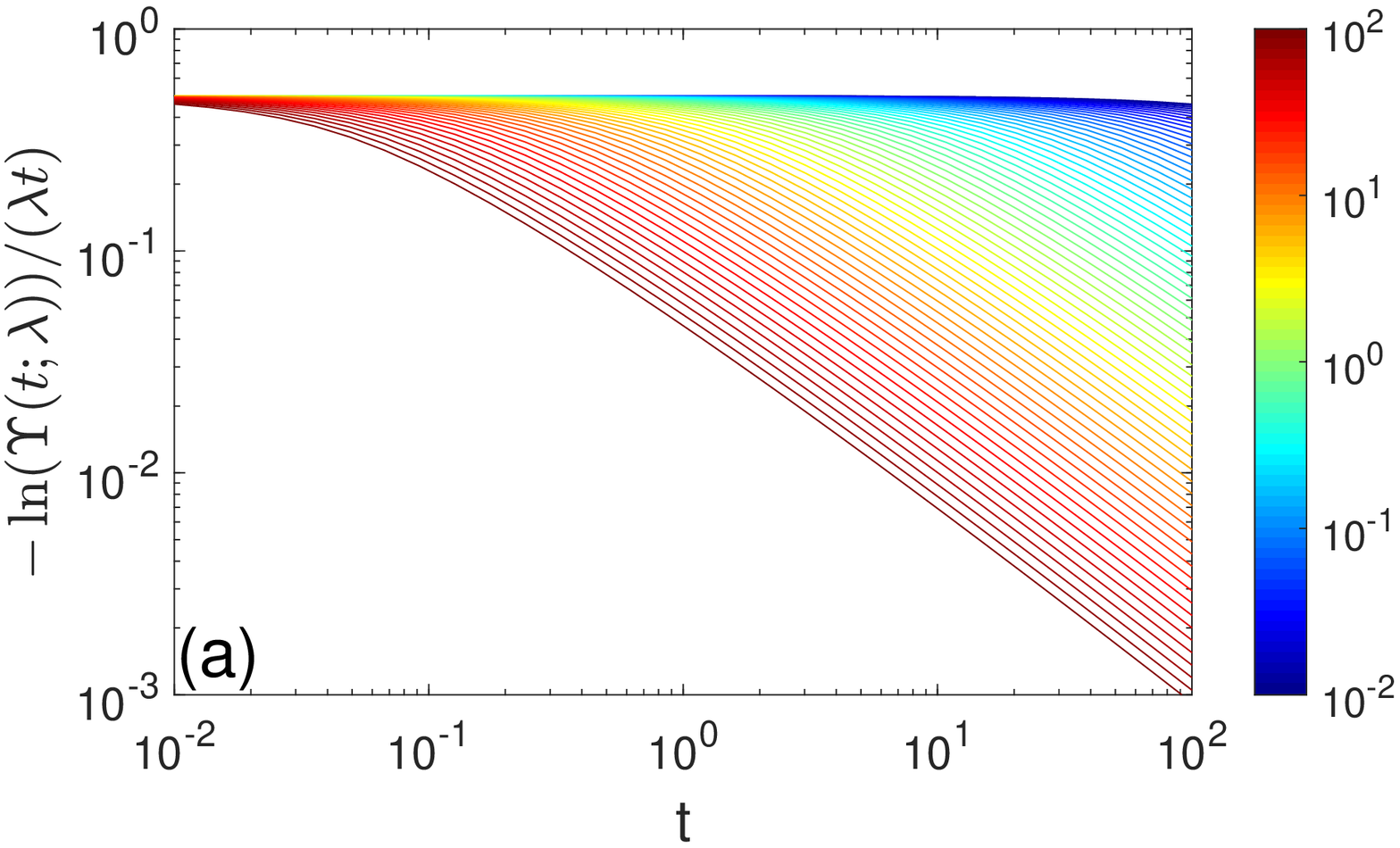}
\includegraphics[width=70mm]{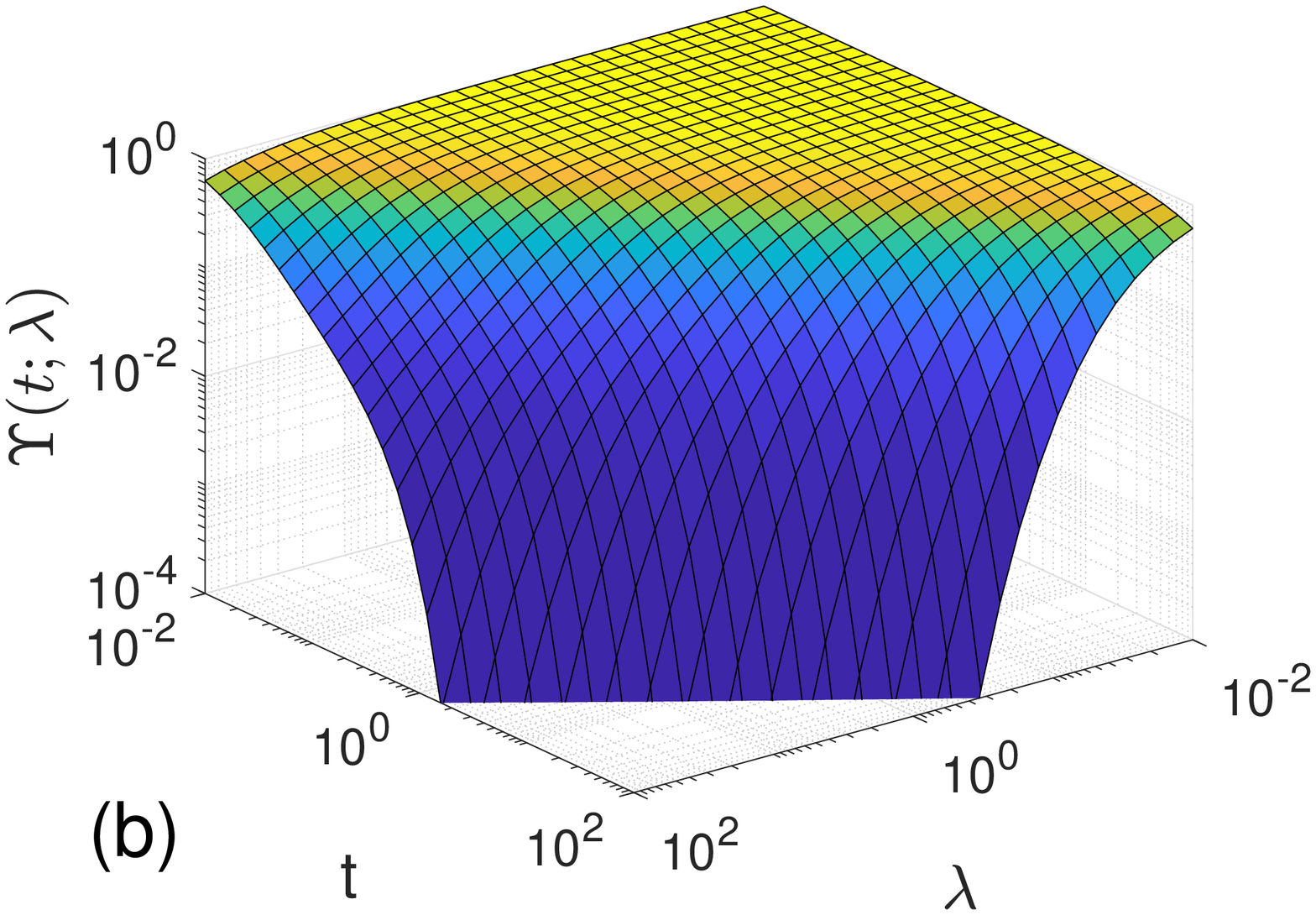}
	\includegraphics[width=82mm]{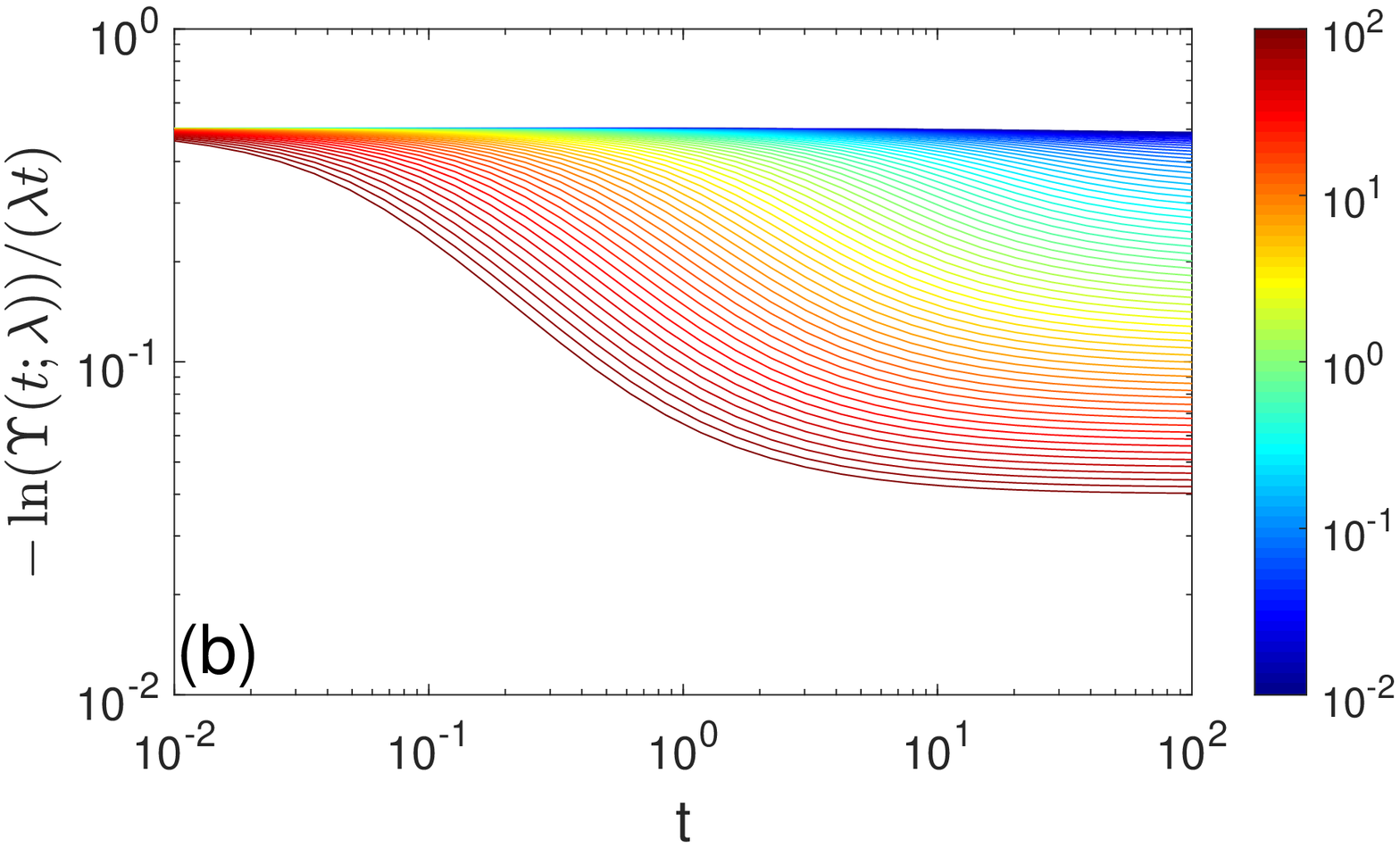}
\includegraphics[width=70mm]{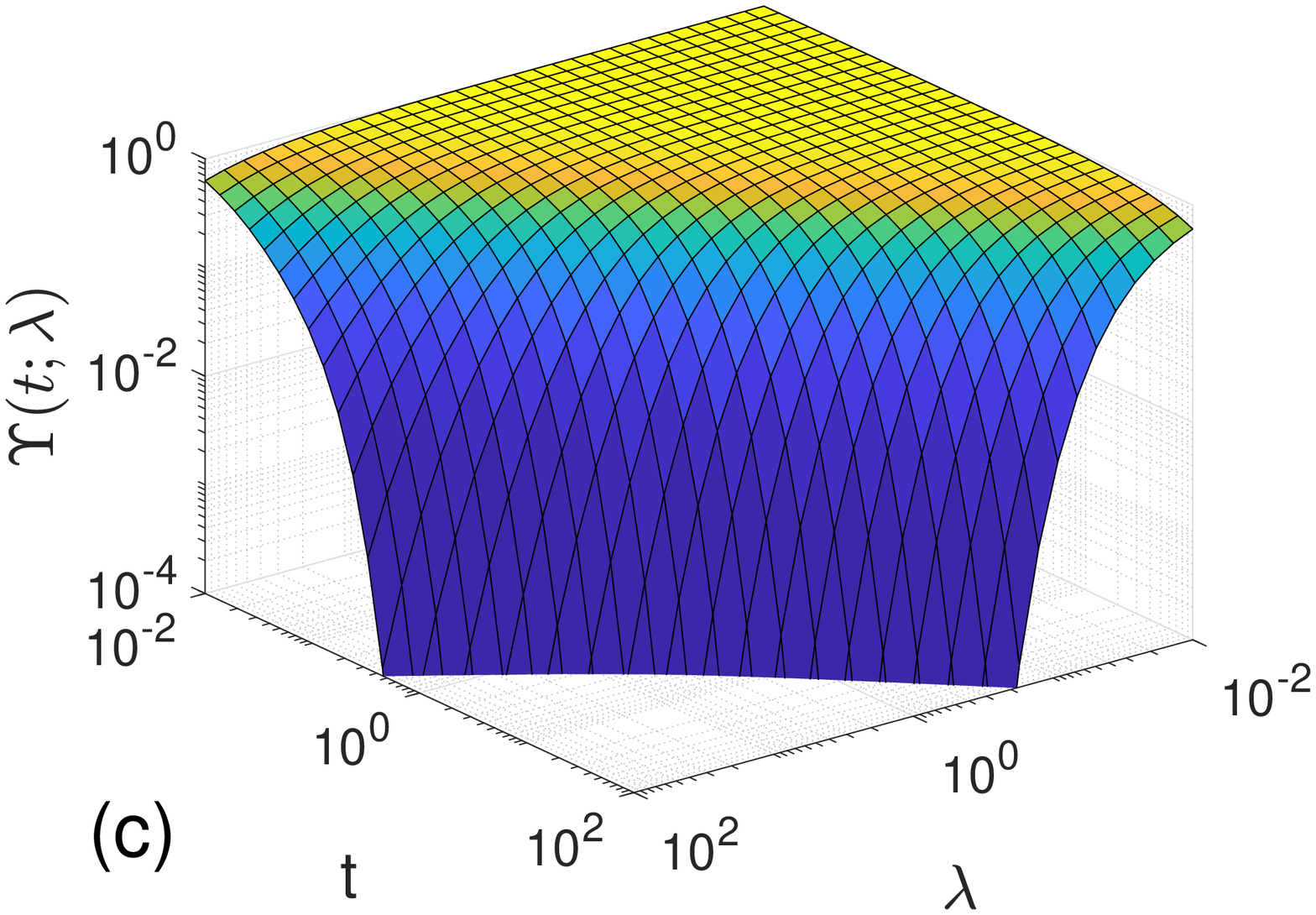}
	\includegraphics[width=82mm]{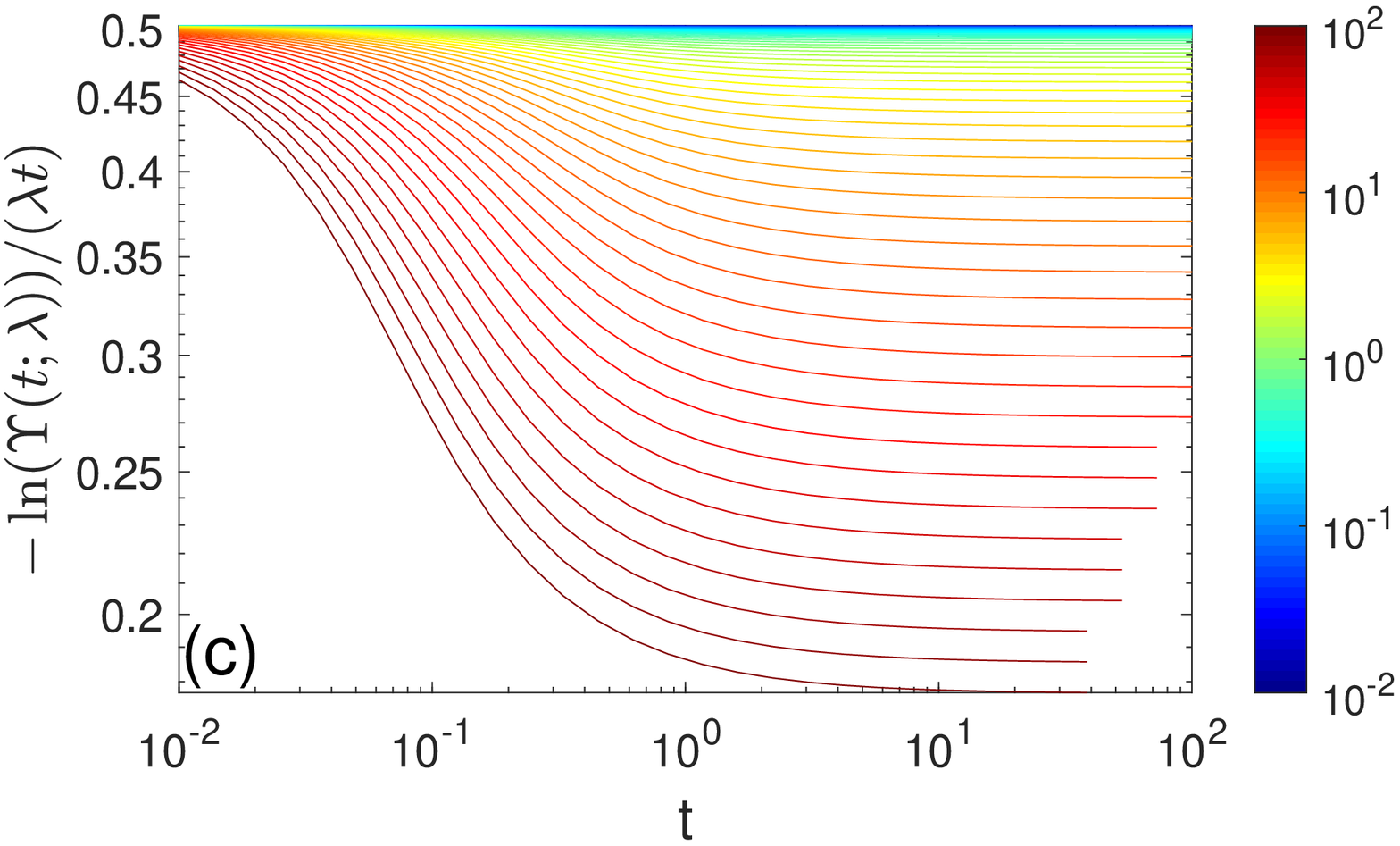}
\end{center}
\caption{
{\bf (Left panels)} The function $\Upsilon(t;\lambda)$ for the
diffusing diffusivity modeled by reflected Brownian motion on
$(0,D_m)$, with $D_m = 1$ and $\sigma = 0$ {\bf (a)}, $\sigma = 0.1$
{\bf (b)}, and $\sigma = 1$ {\bf (c)} (arbitrary units).  For $\sigma
= 0$, Eq. (\ref{eq:Ups_sigma0}) was used, whereas for $\sigma > 0$,
$\Upsilon(t;\lambda)$ was computed from Eq. (\ref{eq:Y_switching}) for
an approximate switching diffusion model with $\dD = 0.01$.  We
checked that a smaller value of $\dD$ yielded similar results (not
shown).  The vertical axis is truncated at $10^{-4}$.  {\bf (Right
panels)} The ratio $-\ln(\Upsilon(t;\lambda))/(t\lambda)$ as a
function of $t$ for the same model and parameters.  Each of 64 curves
corresponds to a value $\lambda$ sampled between $10^{-2}$ and $10^2$
at logarithmic scale (see colorbar).  }
\label{fig:Upsilon_uni}
\end{figure}

Left panels of Fig. \ref{fig:Upsilon_uni} show the behavior of the
function $\Upsilon(t;\lambda)$ for a diffusing diffusivity modeled by
reflected Brownian motion on $(0,D_m)$.  Although the exact solution
is provided in Sec. \ref{sec:rBm}, it is much faster and easier to use
the approximate solution via the switching diffusion model.  As
expected, the function $\Upsilon(t;\lambda)$ approaches $1$ as $t \to
0$ or $\lambda \to 0$.  In turn, when either of these variables is
getting large, $\Upsilon(t;\lambda)$ decreases.  According to
Eq. (\ref{eq:Ups_sigma0}), the decay is slow for the case without
diffusivity dynamics ($\sigma = 0$), see
Fig. \ref{fig:Upsilon_uni}(a).  In turn, much faster decay is observed
for other cases with $\sigma > 0$, in agreement with the asymptotic
analysis of Sec. \ref{sec:rBm}.

To get a closer look into the behavior of $\Upsilon(t;\lambda)$, it is
convenient to plot $-\ln (\Upsilon(t;\lambda))/(t\lambda)$ as a
function of $t$.  At small $t$, one has $T_t \approx D_0 t$ so that
$\Upsilon(t;\lambda) \simeq \langle e^{- t\lambda D_0}\rangle \simeq 1
- t\lambda \langle D_0 \rangle$, where $\langle D_0\rangle$ is the
mean initial diffusivity, which is equal to $D_m/2$ in this model.  As
a consequence, the ratio $-\ln (\Upsilon(t;\lambda))/(t\lambda)$
approaches $D_m/2$ as $t\to 0$.  The opposite limit $t\to \infty$ is
less universal: for instance, Eq. (\ref{eq:Ups_sigma0}) exhibits
$\ln(t)/t$ decay, whereas Eq. (\ref{eq:Ups_uniform}) leads to a
constant.  Right panels of Fig. \ref{fig:Upsilon_uni} illustrate this
behavior.  All shown curves start from the mean diffusivity $D_m/2 =
0.5$ at $t = 0$.  As $t$ grows, all curves decrease but the speed of
decrease depends on $\lambda$ and $\sigma$.  For $\sigma = 0$, the
ratio vanishes as $t\to\infty$ whereas it reaches a nonzero limit
$\sigma^2 \gamma_0(\lambda)/(2\lambda)$ for $\sigma > 0$, where
$\gamma_0(\lambda)$ is the smallest eigenvalue of the operator
$\partial^2_{D_0} - (2\lambda/\sigma^2) D_0$, see Sec. \ref{sec:rBm}.
As a consequence, this limit changes from $D_m/2$ for small $\lambda$
to $|a'_0| (\sigma^2/2)^{1/3} \lambda^{-1/3}$ for large $\lambda$.
One can see that the range of variations of $-\ln
(\Upsilon(t;\lambda))/(t\lambda)$ is getting narrower as $\sigma$
increases.  Indeed, strong fluctuations rapidly mix all diffusivities
in $(0,D_m)$, restoring the behavior with the mean diffusivity
$D_m/2$, as at short times.

\section{Conclusion}

We formulated a unifying approach for studying first-passage time
distributions and related diffusion-limited reactions in the realm of
diffusing diffusivity and switching diffusion models.  In both cases,
the dynamics of randomly changing diffusivity $D_t$ is assumed to be
independent from the particle's position, so that the subordination
argument yields a general spectral expansion for the propagator and
the first-passage time probability density.  The key element coupling
the stochastic diffusivity to the motion of the particle is the
moment-generating function $\Upsilon(t;\lambda|D_0)$ of the integrated
diffusivity.

In diffusing diffusivity models, continuous changes of $D_t$ are
governed by a stochastic differential equation, and the function
$\Upsilon(t;\lambda|D_0)$ can be calculated by using the Feynman-Kac
formula.  We illustrated this formalism for the case when the
diffusivity is modeled by reflected Brownian motion on an interval
with reflecting endpoints.
In turn, when the diffusivity randomly switches between discrete
values, we derived a matrix representation of the function
$\Upsilon(t;\lambda|D_0)$ involving the matrix of switching rates.  We
also formalized the connection between these two classes of models by
relating the coefficients of the stochastic differential equation,
$\mu(D)$ and $\sigma(D)$, to the switching rates.  With the help of
this formalism, one can extend former results on diffusive search
problems
\cite{Loverdo08,Benichou14,Redner,Metzler,Holcman13,Holcman14} to
heterogeneous diffusion, compute the related first-passage time
distributions \cite{Benichou10,Godec16a,Grebenkov18c,Grebenkov18d} and
investigate diffusion-limited reactions in heterogeneous media.

While continuously changing diffusivity may represent the effect of
rapidly re-arranging medium onto the motion of particles, discrete
changes of the diffusivity can mimic switching between conformational
states of a polymer or reversible binding of the diffusing molecule to
other constituents (static or mobile) of the medium.  In particular,
the state with zero diffusivity can incorporate trapping events.
While former studies involving stochastic diffusivity were focused on
a specific choice of the Feller process (which includes as a
particular case the square of the Ornstein-Uhlenbeck process used in
\cite{Jain16,Jain16b,Chechkin17}), the general formalism of the present
paper opens the door to study a very broad class of various processes
in a unified way.  As the microscopic theory expressing the impact of
rapidly re-arranging media onto the particle's dynamics in terms of an
appropriate diffusing diffusivity model is still missing, the
possibility of dealing with a broad range of ``candidate processes''
is particularly valuable for future research.

\end{document}